%% file: main.tex
\documentclass[sigconf]{acmart}

\usepackage[normalem]{ulem}
\useunder{\uline}{\ul}{}
\usepackage{bm}
\usepackage{amsmath}
\usepackage{xfakebold}

\usepackage[linesnumbered,ruled,vlined]{algorithm2e}
\usepackage{xcolor}

\SetCommentSty{mycommfont}

\usepackage{setspace}
\SetKwInput{KwInput}{Input\quad}                % Set the Input
\SetKwInput{KwOutput}{Output~}              % set the Output
\usepackage{varwidth}% http://ctan.org/pkg/varwidth

\usepackage{booktabs, tabularx} 
\definecolor{egyptianblue}{rgb}{0.06, 0.2, 0.65}

\usepackage{comment, array, bm, pifont, mathtools, subcaption, float, graphicx, xcolor, booktabs, xfrac, scalerel, enumerate, enumitem, multirow, pdfrender, placeins, xcolor}

\newcolumntype{M}[1]{>{\centering\arraybackslash}m{#1}}
\newcolumntype{L}[1]{>{\raggedright\arraybackslash}m{#1}}
\newcommand{\xmark}{\ding{55}}
\newcommand*{\boldcheckmark}{%
  \textpdfrender{
    TextRenderingMode=FillStroke,
    LineWidth=.5pt, % half of the line width is outside the normal glyph
  }{\checkmark}%
}

\newcommand{\altfrac}[2]{\ifmmode\def\tmp{$}\else\def\tmp{}\fi\mbox{%
    {\raisebox{.24\ht\strutbox}{\tmp#1\tmp}}%
    \kern-2.2pt\scalebox{1.6}[1.5]{/}\kern-1.8pt%
    {\tmp#2\tmp}%
    }}

\definecolor{ao(english)}{rgb}{0.0, 0.5, 0.0}
\definecolor{cadmiumred}{rgb}{0.89, 0.0, 0.13}
\newcommand{\fbseries}{\unskip\setBold\aftergroup\unsetBold\aftergroup\ignorespaces}
\makeatletter
\newcommand{\setBoldness}[1]{\def\fake@bold{#1}}
\makeatother

\newlength{\Oldarrayrulewidth}

\AtBeginDocument{%
  \providecommand\BibTeX{{%
    \normalfont B\kern-0.5em{\scshape i\kern-0.25em b}\kern-0.8em\TeX}}}

\setcopyright{acmcopyright}

\copyrightyear{2021}
\acmYear{2021}
\setcopyright{rightsretained}
\acmConference[MM '21]{Proceedings of the 29th ACM International
Conference on Multimedia}{October 20--24, 2021}{Virtual Event, China}
\acmBooktitle{Proceedings of the 29th ACM International Conference on
Multimedia (MM '21), October 20--24, 2021, Virtual Event,
China}
\acmISBN{978-1-4503-8651-7/21/10}
\acmDOI{10.1145/3474085.3475320}

\acmSubmissionID{864}

\begin{document}

% \title{Meta Transfer Learning for Video Super-Resolution}
\title{Ada-VSR: Adaptive Video Super-Resolution with Meta-Learning}

\author{Akash Gupta, ~Padmaja Jonnalagedda, ~Bir Bhanu, ~Amit K. Roy-Chowdhury}
\orcid{1234-5678-9012}
\affiliation{%
  \institution{University of California, Riverside \\ $\left\{\text{agupt013@, sjonn002@, bir.bhanu@, amitrc@ece}\right\}$.ucr.edu}
% %   \streetaddress{900 University Ave}
% %   \city{Riverside}
% %   \state{California}
  \country{}
% %   \postcode{92521}
}
% %\authornote{Corresponding Author}
% \email{{agupt013@, sjonn002@, bir.bhanu@, amitrc@ece}.ucr.edu}
% \email{\begin{varwidth}{1mm}\centering\end{varwidth}}

%  \email{amitrc@ece.ucr.edu}

% %%
% %% By default, the full list of authors will be used in the page
% %% headers. Often, this list is too long, and will overlap
% %% other information printed in the page headers. This command allows
% %% the author to define a more concise list
% %% of authors' names for this purpose.

\begin{abstract}
    \input{sections/0_abstract}
\end{abstract}

\begin{CCSXML}
<ccs2012>
   <concept>
       <concept_id>10010147.10010257.10010293.10010315</concept_id>
       <concept_desc>Computing methodologies~Instance-based learning</concept_desc>
       <concept_significance>300</concept_significance>
       </concept>
   <concept>
       <concept_id>10010147.10010257.10010258.10010262.10010277</concept_id>
       <concept_desc>Computing methodologies~Transfer learning</concept_desc>
       <concept_significance>300</concept_significance>
       </concept>
   <concept>
       <concept_id>10010147.10010178.10010224.10010245.10010254</concept_id>
       <concept_desc>Computing methodologies~Reconstruction</concept_desc>
       <concept_significance>300</concept_significance>
       </concept>
 </ccs2012>
\end{CCSXML}

\ccsdesc[300]{Computing methodologies~Instance-based learning}
\ccsdesc[300]{Computing methodologies~Transfer learning}
\ccsdesc[300]{Computing methodologies~Reconstruction}
\keywords{Video Super-resolution, Temporal Super-resolution, Meta-Transfer learning, Internal Learning}


\maketitle

%%%%%%%%%%%%%%%%%%%%%%%%%%%%%%%%%%%%%%%%%%%%%%%%%%%%%%%%%%%%%%%%%%%%

%%
%% Introduction
\input{sections/1_introduction}

%%
%% Related work
\input{sections/2_related_work}

%%
%% Problem Formulation
%\input{sections/3_problem_formulation}

%%
%% Approach
\input{sections/3_approach}

%%
%% Experiments
\input{sections/4_experiments}

%%
%% Conclusion
\input{sections/5_conclusion}

%%
%% Conclusion
\input{sections/6_acknowledgments}


\balance
\bibliographystyle{ACM-Reference-Format}
\bibliography{sample-base}

\end{document}

%% file: sections/0_abstract.tex
Most of the existing works in supervised spatio-temporal video super-resolution (STVSR) heavily rely on a large-scale external dataset consisting of paired low-resolution low-frame rate (LR-LFR) and high-resolution high-frame rate (HR-HFR) videos. Despite their remarkable performance, these methods make a prior assumption that the low-resolution video is obtained by down-scaling the high-resolution video using a known degradation kernel, which does not hold in practical settings. Another problem with these methods is that they cannot exploit instance-specific internal information of a video at testing time. Recently, deep internal learning approaches have gained attention due to their ability to utilize the instance-specific statistics of a video. However, these methods have a large inference time as they require thousands of gradient updates to learn the intrinsic structure of the data.
 In this work, we present \textbf{Ada}ptive \textbf{V}ideo \textbf{S}uper-\textbf{R}esolution (\textbf{Ada-VSR}) which leverages external, as well as internal, information through meta-transfer learning and internal learning, respectively. 
 Specifically, meta-learning is employed to obtain adaptive parameters, using a large-scale external dataset, that can adapt quickly to the novel condition (degradation model) of the given test video during the internal learning task, thereby exploiting external and internal information of a video for super-resolution.
 The model trained using our approach can quickly adapt to a specific video condition with only a few gradient updates, which reduces the inference time significantly. Extensive experiments on standard datasets demonstrate that our method performs favorably against various state-of-the-art approaches. The project page is available at \texttt{\url{https://agupt013.github.io/AdaVSR.html}}

%% file: sections/1_introduction.tex
\section{Introduction}

\begin{figure}
    \centering
    \includegraphics[width=0.98\columnwidth]{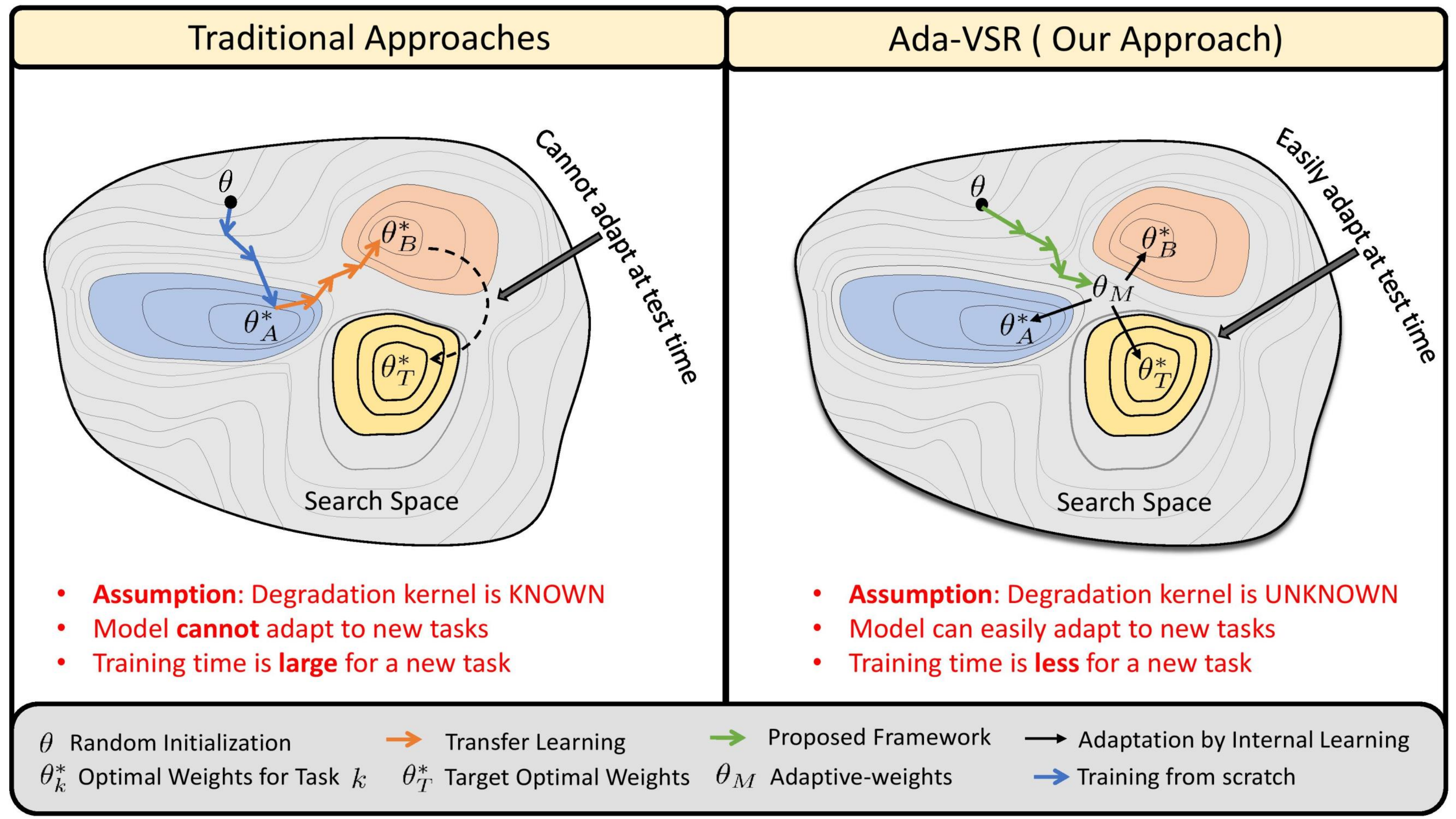}
    % \vspace{-2mm}
    \caption{Comparison of traditional approaches against Ada-VSR for the blind VSR task. Traditional supervised approaches train their model assuming the degradation kernel for task A is known (left; blue arrows). Transfer learning can be adopted to find optimal model parameters for Task B with a different degradation kernel (left; orange arrows). However, the model will not be able to generalize for the target task T when degradation is not known. On the other hand, our proposed approach tries to find weights that can easily adapt to the target task with only a few gradient updates via internal learning (right; green arrows. See sec.~\ref{ssec:int_learn})}
    \vspace{-2mm}
    % for more details).}
    \label{fig:teaser}
    
\end{figure}

With the increasing popularity of high-performance higher resolution displays such as 4K Ultra HD (UHD), the demand for high-quality visual content is also increasing. 
However, professional video production and TV screen content are still at Full HD (1080p) resolution~\cite{srvfr, kim2020fisr, kim2019deep}. 
As rendering low-resolution content on higher resolution displays lowers perceptual quality, it calls for improving the resolution of the content to match that of the display.
Enhancing the quality of video not only requires increasing the spatial resolution but also the temporal resolution for smooth rendering on high-performance displays. 
Therefore, it is critical to improve the spatial as well as the temporal resolution of videos to enhance the perceptual quality.
Similarly, in microscopy time-lapse imagery, the quality of imaging plays a crucial role in determining the performance of the cell tracking algorithm~\cite{comes2019influence, ferri2018time, su2021spatio}. Increasing the spatial and temporal resolution of the time-lapse video is desirable to facilitate automatic cell detection and accurate reconstruction of the cell trajectories, respectively. However, capturing the time-lapse video with higher spatio-temporal resolution damages the cell and can lead to disposal of the painstakingly collected experimental data~\cite{comes2019influence, ferri2018time, gupta2020deep}.

Most existing approaches have addressed the task of video spatial super-resolution (VSR)~\cite{kappeler2016video, jo2018deep, xue2019video, tao2017detail, caballero2017real, wang2019edvr} and temporal video super-resolution (TSR)~\cite{mahajan2009moving, zitnick2004high, bao2019depth, bao2019memc, jiang2018super, liu2017video, gupta2020alanet}, separately. 
A straightforward strategy to perform spatio-temporal video super-resolution (STVSR) is to cascade the VSR model and the TSR model to generate high-resolution high-frame rate (HR-HFR) video from low-resolution low-frame rate (LR-LFR) video. Nevertheless, this does not yield optimal results as it cannot fully utilize the available spatio-temporal information~\cite{xiao2020space}. 
Recently, a few works~\cite{kim2019deep, kim2020fisr, xiao2020space, xiang2020zooming}, studied the problem of joint spatio-temporal video super-resolution. 
Zooming Slow-Mo~\cite{xiang2020zooming} proposed a one-stage STVSR framework using Deformable Convolutional LSTM. The authors in~\cite{xiao2020space} utilize temporal profiles to exploit spatio-temporal information. FISR~\cite{kim2020fisr} proposes a multi-scale temporal loss for joint frame-interpolation and super-resolution. 
However, these approaches require a large dataset of LR-HR pairs with the assumption that the down-sampling kernel to obtain LR frames from HR frames is known and fixed, which does not hold true in a real world setting (Figure~\ref{fig:teaser}, left). 
The problem of blind SR in images, where down-sampling kernel is unknown, is tackled either by estimating the down-sampling kernel~\cite{gu2019blind, bell2019blind} or by exploiting the deep internal prior~\cite{shocher2018zero, ulyanov2018deep} to learn the internal structure of the image. Consequently, these approaches achieve good performance at the expense of heavy computational time as it requires thousands of back-propagation gradient updates for each instance. Another shortcoming of such approaches is that they cannot take advantage of a pre-trained network learned using a large-scale external dataset~\cite{soh2020meta}.

Meta-learning has recently garnered much interest to tackle the aforementioned shortcomings. Meta-learning aims to adapt quickly and efficiently to unseen data available at inference time. There are three common approaches to meta-learning: metric-based~\cite{snell2017prototypical, sung2018learning, vinyals2016matching}, model-based~\cite{santoro2016meta, oreshkin2018tadam, mishra2017simple}, and optimization-based~\cite{grant2018recasting, finn2017meta, finn2017model}. 
Model-Agnostic Meta-Learning
(MAML)~\cite{finn2017model} is a gradient-based method and has shown impressive performance by learning the optimal initial state of the model such that it can quickly adapt to a new task with a few gradient steps.

In this paper, we introduce a novel framework \textbf{Ada}ptive \textbf{V}ideo \textbf{S}uper-\textbf{R}esolution (\textbf{Ada-VSR}) which aims to generate high resolution high-frame rate (HR-HFR) video from a low-resolution low-frame rate (LR-LFR) input. Inspired by meta-transfer learning, we utilize external knowledge as well as internal knowledge from videos for the task of joint spatio-temporal video super-resolution (STVSR). Our approach leverages knowledge from the external dataset and learns adaptive model parameters using meta-learning.
% To this end, we first pre-train a model using a large-scale external dataset containing LR-LFR frames obtained by bi-cubic down-sampling of HR-HFR frames. 
As shown in Figure~\ref{fig:teaser} (right), meta-learning is employed to learn initial model parameters that can quickly and efficiently adapt to the test video with unknown degradation. 
% \edits{Akash: For external and Internal Learning.}
% \amitnew{Are you going to add this? Do not use "right" space. It is not clear what it means. You can say something like "adaptation to happen more efficiently for fast learning."}
Specifically, we use different down-sampling kernels on a large-scale dataset, as an external learning task, to learn a model that is easy and fast to finetune for novel tasks. The model parameters obtained using external learning allow adaptation to happen more efficiently for fast learning. Next, internal learning is leveraged to finetune the initial model to learn video-instance specific knowledge with limited gradient steps. Since there are only a few gradient steps involved during internal learning for each video, our approach is significantly faster when compared to approaches that completely rely on internal learning and require thousands of gradient updates.\vspace{-2mm}

\subsection{Approach Overview}
\begin{figure*}[t]
    \centering
    \subfloat[External Learning involves two steps: task-specific training (left) and blind task adaptation (right). Using external dataset  $\textbf{V}_{HR}\in \mathcal{D}_{HR}$ we create a meta-batch with dynamic task generator by spatial down-scaling using $f_s$ and temporal down-scaling using $f_t$. The meta-batch consists of a train and a test set.  First, the train set is used to learn optimal model for task specific training. Then the learnt model is adapted to the blind test task. Following this meta-learning strategy, we obtain parameters which can quickly adapt model during internal learning.]{\includegraphics[width=0.98\linewidth]{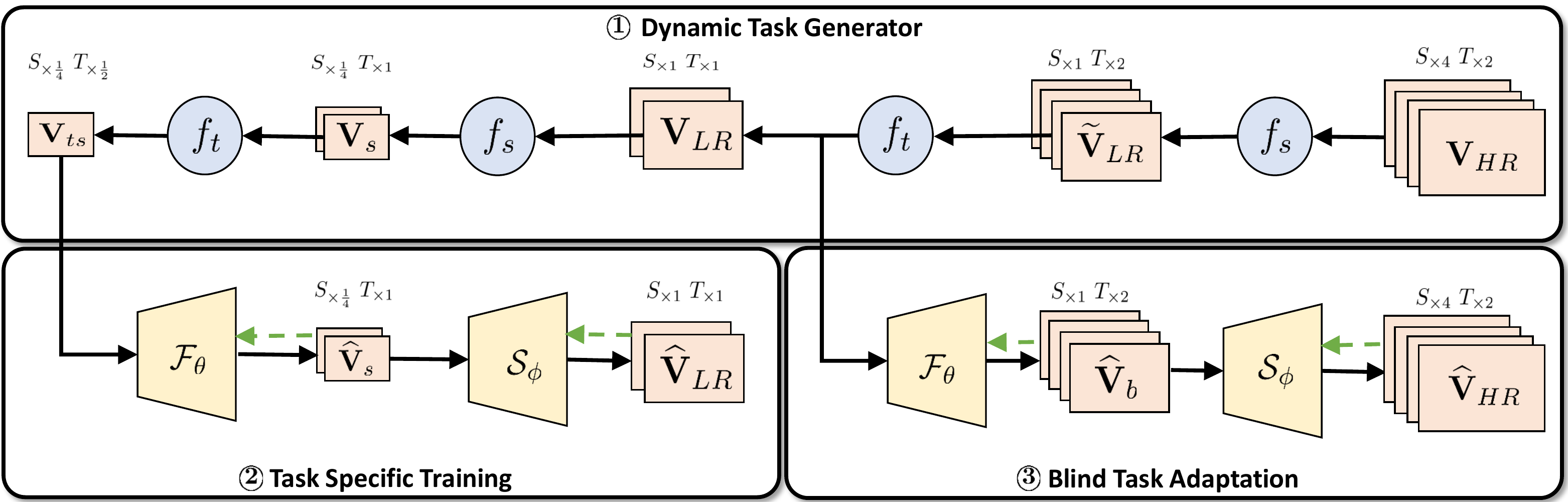}}
    \hfill
    \subfloat[Internal Learning and Inference. Left: We exploit the instance specific information during internal learning. The LR-LFR test video $\textbf{V}_{LR}$ is used to obtain a spatio-temporal down-sampled video ($\textbf{V}_{I}$). For internal learning, the down-sampled video $\textbf{V}_{I}$ is used to reconstruct $\widehat{\textbf{V}}_{LR}$. This helps network to learn the internal statistic of the LR-LFR test video. Right: Finally, HR-HFR video $\widehat{\textbf{V}}_{HR}$ is generated by passing the test video $\textbf{V}_{LR}$ through the model trained via internal-learning. ]{\includegraphics[width=0.98\linewidth]{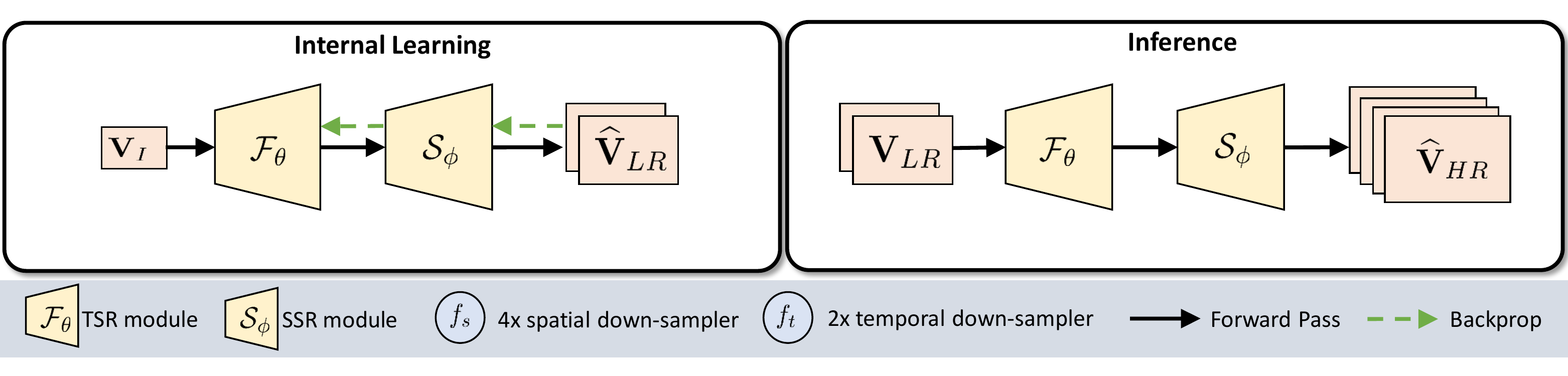}}
    \vspace{-2mm}
    \caption{Overview of Adaptive Video Super-Resolution (Ada-VSR) framework. Our framework consists of two modules External Learning and Internal Learning. (a) External learning leverages meta-training protocol and exploits the external dataset to learn parameters that can easily adapt to novel tasks (different degradation in our case). (b) Internal learning helps exploit internal structure of the given video and is used to generate a HF-HFR video ($\widehat{\textbf{V}}_{HR}$) from LR-LFR video $\textbf{V}_{LR}$. Since internal learning is initialized by the adaptive parameters obtained through external dataset, our model can quickly adapt to video degraded by unknown kernel with only a few gradient steps. 
    %\amit{It is not clear what the input and output videos area, what is the external dataset. What is the green dotted arrow? It is hard to follow the flow, especially in the top figure.}\edits{Akash: Updated the figure and added numbers for each step. Variables with hat are the output of the networks. Dynamic task generator is generating data that is used to train the models.} 
    % \amitnew{Caption needs to be updated to address the comment. What is the input in the figure - it is not clear.}
    }
    \label{fig:overview}
\end{figure*}

An overview of our Adaptive Video Super-Resolution (\textbf{Ada-VSR}) training scheme is illustrated in Figure~\ref{fig:overview}. Given a low-resolution low frame-rate input, our objective is to generate a high resolution high frame-rate video when the degradation kernel is unknown. Our \textbf{Ada-VSR} approach consists of two networks: the temporal super-resolution module (TSR) denoted by $\mathcal{F}_{\theta}$ and the spatial super-resolution module (SSR) as $\mathcal{S}_{\phi}$. We adopt a meta-learning framework to train both the networks jointly using an external dataset containing LR-LFR (obtained using dynamic task generator; refer Fig.~\ref{fig:overview}) and HR-HFR video pairs. The objective of the meta-training is to learn model parameters that can be easily adapted to the test video. This meta-training is performed only once and the adaptive parameters are used to initialize the model in the next step. In a practical setting, we will only have access to the LR-LFR video. Hence, we exploit the internal structure within the test LR-LFR video using internal learning. The objective of internal learning is to finetune the model for the video instance to improve the spatio-temporal super-resolution with only a few gradient steps. Therefore, we first downscale the LR-LFR further to obtain a super low-resolution low-frame rate (SLR-LFR) video. Secondly, we finetune the model to reconstruct the LR-LFR video from the obtained SLR-LFR video to utilize the instance-specific knowledge. As the parameters obtained by external-learning are learnt for fast learning, internal learning allows quick adaptation of the model parameters to blind spatio-temporal super-resolution task.
%\amitnew{I do not like the term right space. Please change it.}
% Finally, the model trained with internal learning is used to infer HF-HFR video from LF-LFR video.
Finally, the trained model is used to infer HF-HFR video from LF-LFR video.

% In the meta-training, we dynamically generate a batch of LR-LFR and HR-HFR pair by down-sampling it spatially and temporally using different blur kernels to obtain different tasks. Each combination of a specific spatial and a specific temporal down-sampling is considered a separate task (see section~\ref{}).  , to train the TSR and SSR model aim to leverage the external dataset $\mathcal{D}_{ext}$ to learn a warm initialization for the meta-training of our model. We create a synthetic dataset of by selecting alternate frame from a video and applying bi-cubic down-sampling to obtain a LR-LFR and HR-HFR paired data. We use this synthetic data pairs to pre-training.  consisting of high-resolution high-frame rate video, we aim first pre-training a For a given model $f$ parameters by $\theta$ 

\subsection{Contributions}
The key contributions of our proposed framework are as follows.
%summarized
\begin{itemize}[leftmargin=*]
    \setlength \itemsep{0.4em}
    \item We propose a novel meta-learning framework \textbf{Ada-VSR} for the task of joint spatio-temporal super-resolution by leveraging external and internal learning.
    \item We employ a combination of various spatial and temporal down-sampling techniques during training to learn a model that can easily adapt to unknown down-sampling/degradation process.
    \item We significantly reduce the computational time by greatly reducing the gradient steps required during internal learning. 
\end{itemize}

%% file: sections/2_related_work.tex
\section{Related Work}
\input{tables/related_work}

Our work relates to research in spatial and temporal video super-resolution, internal learning and, meta-transfer learning. In this section, we discuss some methods closely related to our work. We provide a characteristic comparison of recent works in Table~\ref{tab:compare_methods}.

\noindent \textbf{Image Super-Resolution.} Deep learning approaches have shown remarkable performance on the task of image super-resolution~\cite{haris2018deep, dong2015image, kim2016accurate, ledig2017photo, lim2017enhanced, shocher2018zero, gu2019blind, zhang2018learning}. Recently, various Convolutional Neural Network (CNN) based approaches have been proposed for non-blind image SR where the down-sampling kernel (e.g. bicubic), used to obtain low-resolution (LR) image, is known~\cite{haris2018deep, dong2015image, kim2016accurate, ledig2017photo, lim2017enhanced}. Despite the impressive performance of these methods, their efficacy deteriorates when the down-sampling kernel is different than the one used to train these models due to the domain gap. To overcome this issue, SRMD~\cite{zhang2018learning} incorporates multiple degradation kernels as input to their model along with the LR image. On the other hand, Zero-Shot Super-Resolution (ZSSR)~\cite{shocher2018zero} exploits the deep prior~\cite{ulyanov2018deep} to learn an image specific structure to obtain an SR image.
Some approaches first try to estimate the degradation kernel and utilize the estimated kernel for image super-resolution. An iterative approach to correct inaccurate degradation kernels is introduced in~\cite{gu2019blind}. Similar to ZSSR, the KernelGAN~\cite{bell2019blind} utilizes the patch-recurrence property of a single image for super-resolution. However, these methods train the network from scratch for all the image instances, making them computationally heavy.\\

\noindent \textbf{Video Spatial Super-Resolution.} 
Earlier works in video super-resolution focused on developing effective priors on the HR frames to solve this problem~\cite{bascle1996motion, cain2001projection, farsiu2004fast, shahar2011space}. Motivated by the success of deep learning approaches in image super-resolution~\cite{shocher2018zero, zhang2018learning, gu2019blind}, several deep learning based methods have been proposed for video super-resolution ~\cite{kappeler2016video, jo2018deep, xue2019video}. A CNN based approach is proposed in~\cite{kappeler2016video}, where the network is trained on both the spatial and temporal dimensions of videos to spatially enhance the frames. An SR draft-ensemble approach for fast video spatial super-resolution is proposed in~\cite{liao2015video}.  In~\cite{tao2017detail, caballero2017real}, the authors incorporate optical flow estimation models to explicitly account for the motion between neighboring frames. However, accurate flow is difficult to obtain given occlusion and large motions. A computationally lighter flow estimation module (TOFlow) is proposed in~\cite{xue2019video} to account for motion information. DUF~\cite{jo2018deep} overcomes this problem by implicit motion compensation using their proposed dynamic upsampling filter network. Pyramid, Cascading and Deformable convolution (PCD) alignment and the Temporal and Spatial Attention (TSA) modules are proposed in EDVR~\cite{wang2019edvr} to incorporate implicit motion compensation. However, these approaches assume the degradation kernel for down-sampling is known and/or require a large amount LR-HR pairs to train their models.\smallskip

\noindent \textbf{Video Temporal Super-Resolution.}  Video super-resolution can also be performed in the temporal dimension and is often termed as video interpolation. In temporal video super-resolution, the task is to generate a high-frame rate (HFR) video from a low-frame rate video (LFR). Existing approaches \cite{mahajan2009moving, zitnick2004high, bao2019depth, bao2019memc, jiang2018super, liu2017video} use optical flow estimation between input frames for temporal super-resolution. Thus, the quality of estimated optical flow governs the quality of frame interpolation. Deep learning approaches have demonstrated  effectiveness in temporal super-resolution tasks. A straightforward application of CNNs for intermediate frame synthesis is presented in~\cite{long2016learning}. Some methods \cite{niklaus2017video, niklaus2017videosepconv} apply CNNs to estimate space-varying and separable convolutional kernels for frame synthesis using neighbourhood pixels. \cite{aich2020non} proposed a non-adversarial approach to generate videos by first learning optimized representation and then interpolating between the optimized latent representation of two frames to synthesize central frame. Joint video deblurring and interpolation to enhance and increase the frame-rate of a video is explored in~\cite{shen2020blurry, gupta2020alanet}. These approaches do not perform spatial super-resolution in their work and assume that the low-temporal resolution video is obtained by averaging 9 consecutive frames. Unlike these methods, we address the task of joint spatial and temporal super-resolution. \smallskip

\noindent\textbf{Meta-Learning.}
Recently, meta-learning algorithms have achieved impressive performance in various applications like few-shot learning~\cite{li2017meta, jamal2019task, ren2018meta, sun2019meta, snell2017prototypical}, reinforcement learning~\cite{schweighofer2003meta, finn2017model, gupta2018unsupervised, nagabandi2018learning} and image super-resolution~\cite{soh2020meta, lee2021dynavsr}. Meta-learning aims to learn a model that can quickly and efficiently adapt to novel unseen tasks. There are three common approaches to meta-learning: metric-based~\cite{snell2017prototypical, sung2018learning, vinyals2016matching}, model-based~\cite{santoro2016meta, oreshkin2018tadam, mishra2017simple}, and optimization-based~\cite{grant2018recasting, finn2017meta, finn2017model}.
DynaVSR is proposed in~\cite{lee2021dynavsr}, which utilizes meta-learning for spatial video super-resolution and has shown superior performance. Different from DynaVSR~\cite{lee2021dynavsr}, we leverage meta-learning for the task of joint spatio-temporal video super-resolution.

%% file: tables/related_work.tex
\begin{table}[b]
\caption{\textit{Categorization of prior works in video super-resolution}. Different from the state-of-the-art approaches, we employ meta-learning to perform blind spatio-temporal video super-resolution. 
% \amit{This table would be more meaningful if, instead of Learning Method, you were able to classify the existing approaches in terms of the assumptions they make, which is indirectly related to the learning method. For example, they may be assuming the existence of paired LR-HR videos. That is more fundamental; the learning method is a tool that can change.}
% \edits{Akash: Updated the table.}
}
\vspace{-1em}
\begin{center}
\resizebox{\columnwidth}{!}{%
\renewcommand{\arraystretch}{1.35}
\begin{tabular}{L{3.1cm}|M{1.15cm}|M{1.15cm}|M{1.15cm}|c}
\toprule[1.2pt]
\multicolumn{1}{c|}{\multirow{2}{*}{Methods}} & \multicolumn{3}{c|}{Super-Resolution} & \multicolumn{1}{c}{{\multirow{2}{*}{Fast Adaptation?}}}\\
\cline{2-4}
& Spatial  & Temporal & Blind  \\
\toprule[0.7pt]
% CNN~\cite{kappeler2016video}         & \textcolor{ao(english)}{\boldcheckmark}   & \textcolor{cadmiumred}{\xmark}            & \textcolor{cadmiumred}{\xmark}          & \textcolor{cadmiumred}{\xmark}\\
\hline
DyaVSR~\cite{lee2021dynavsr}        &  \textcolor{ao(english)}{\boldcheckmark}  & \textcolor{cadmiumred}{\xmark}  &  \textcolor{ao(english)}{\boldcheckmark} & \textcolor{ao(english)}{\boldcheckmark}    \\
\hline
Temporal Profiles~\cite{xiao2020space}       & \textcolor{ao(english)}{\boldcheckmark}   & \textcolor{ao(english)}{\boldcheckmark}   &  \textcolor{cadmiumred}{\xmark }    & \textcolor{cadmiumred}{\xmark } \\
\hline
Zooming Slow-Mo~\cite{xiang2020zooming}       & \textcolor{ao(english)}{\boldcheckmark}   & \textcolor{ao(english)}{\boldcheckmark}   & \textcolor{ao(english)}{\boldcheckmark}   &  \textcolor{cadmiumred}{\xmark }     \\
\hline
\textbf{Ada-VSR (Ours)}           & \textcolor{ao(english)}{\boldcheckmark}   & \textcolor{ao(english)}{\boldcheckmark}   & \textcolor{ao(english)}{\boldcheckmark}   & \textcolor{ao(english)}{\boldcheckmark}\\
\bottomrule[1.2pt]

\end{tabular}
}
\end{center}
\label{tab:compare_methods}

\end{table}

%% file: sections/3_approach.tex
\section{METHODOLOGY}

Given a low-resolution low frame-rate video our goal is to generate a high-resolution high frame-rate video in blind video super-resolution setting where the down-scaling kernel is not known at the test time.
Let the low-resolution low frame-rate video be denoted by
    $ \textbf{V}_{LR} = \begin{bmatrix} ~\mathsf{L}_1, ~\mathsf{L}_2, \cdots,~\mathsf{L}_L\end{bmatrix}$,
with $M$ frames where $\mathsf{L}_t  \in\mathbb{R}^{H \times W \times C}$ and t denotes the time step.
We aim to generate a high-resolution high frame-rate video
    $\textbf{V}_{HR} = \begin{bmatrix} \mathsf{S}_1, ~\mathsf{S}_2, \cdots, ~\mathsf{S}_N\end{bmatrix}$ 
with $N$ frames, where $\mathsf{S}_t \in\mathbb{R}^{\textbf{a}H \times \textbf{a}W \times C}$ and $N = \textbf{b}M$. %
Our objective is to increase the spatial resolution of the given input video $\textbf{V}_{LR}$ by a factor of $\textbf{a}$ and the temporal resolution by a factor of $\textbf{b}$.

%\subsection{Adaptive Video Super-Resolution}

% \begin{figure*}
%     \centering
%     \includegraphics[width=0.98\linewidth]{images/overview.pdf}
%     \vspace{-3mm}
%     \caption{Overview of Adaptive Video Super-Resolution (Ada-VSR) framework.}
    
%     \label{fig:overview}
% \end{figure*}

In this section, we describe the proposed \textbf{Ada-VSR} framework in detail. Our framework consists of two modules: the Temporal Super-Resolution (TSR) module $\mathcal{F}_{\theta}$ and the Spatial Super-Resolution (SSR) module $\mathcal{S}_{\phi}$. We use the TSR module to interpolate frames and increase the frame rate by a factor of 2. The SSR network uses the output of the temporal super-resolution module and increases the spatial resolution by a scaling factor of 4. The scaling factor values are fixed for all our experiments.
% \amit{Are 2 and 4 hard-coded into the method, or can they be generalized? Should specify that.} 
The overall training scheme of our approach $\textbf{Ada-VSR}$ is shown in Figure~\ref{fig:overview}. Our framework consists of two training paradigms: external learning and internal learning.

\subsection{External Learning}
The external learning protocol leverages a large-scale external dataset to perform knowledge transfer and domain generalization using pre-training and meta-transfer learning respectively.\smallskip

\noindent \textbf{Large-Scale Training.}
In large-scale pre-training, we utilize a high-quality external dataset ($\mathcal{D}_{HR}$) to provide a warm start for meta-transfer learning. Since super-resolution tasks with different down-scaling kernels share similar parameter space, large-scale training helps to estimate the natural prior of high-resolution high-frame rate videos. The large-scale pre-training is also effective to stabilize the training of the meta-learning algorithm MAML~\cite{finn2017model}.

For the SSR module, we apply bi-cubic spatial degradation to HR-HFR video $\textbf{V}_{HR} \in \mathcal{D}_{HR} $ 
% \amit{does this belong to $\mathcal{D}_{HR}$?}
to obtain low-resolution high-frame rate video LR-HFR $\widetilde{\textbf{V}}_{LR}$. The videos $\textbf{V}_{HR}$ and $\widetilde{\textbf{V}}_{LR}$ form a synthetic dataset $\mathcal{D}_s$. We train the network $\mathcal{S}_{\phi}$ to learn spatial super-resolution task by minimizing the $\ell_1$ reconstruction loss between all the frames of the generated HR-HFR video ($\widehat{\textbf{V}}_{HR}$) and corresponding ground truth HR-HFR video $\textbf{V}_{HR}$. The objective function for large-scale training of the SSR module $\mathcal{S}_{\phi}$ is defined as:
\begin{align}
    \mathcal{L}^{\mathcal{D}_s} =  \sum\limits_{(\widetilde{\textbf{V}}_{LR}, \textbf{V}_{HR})\sim\mathcal{D}_s}\Big\Vert \mathcal{S}_{\phi}(\widetilde{\textbf{V}}_{LR})-\textbf{V}_{HR}\Big\Vert_1
    \label{eqn:L_Ds}
\end{align}
We choose $\ell_1$-loss instead of Mean-Squared Error (MSE) $\ell_2$ loss as latter has inherent property of generating blurry output as shown in the literature \cite{zhao2015loss}.

% \amit{I am not following the notation here and in Fig. 2.}\edits{Akash: The Large-scale pre-training is not shown in Fig 2. But I have modified the figure for $\widetilde{\textbf{V}}_{LR}$ to be consistent.}

The TSR module should be able to increase the frame-rate of a low-frame-rate (LFR) video. We can interpolate the frames to increase the frame-rate by factor of 2 and learn a residual by minimizing the reconstruction loss between the generated high-frame-rate $\widehat{\textbf{V}}_{HR}$ and the ground truth video $\textbf{V}_{HR}$. However, it may not be able to capture the temporal dynamics efficiently~\cite{xiao2020space, berman2018measuring}.  
Recently some works have addressed this by taking temporal profile 
%or across dimension patches
% \edits{Akash: Same as temporal profile. I removed the term to avoid any confusion.}\amit{what is an "across dimension" patch?} 
to leverage the patch recurrence in temporal dimension to train the network efficiently~\cite{xiao2020space, berman2018measuring}. We adopt the same strategy to train the TSR module $\mathcal{F}_{\theta}$. We define a temporal profile generator function $f_r$ that takes a video input, performs the bi-cubic interpolation in temporal dimension and returns the temporal profile. 
To generate a dataset to train the TSR module we select alternate frames of the high-frame-rate (HFR) video $\textbf{V}_{HR}$ to generate a LFR video $\overline{\textbf{V}}_{LR}$. Then we apply the temporal profile generator function ($f_r$) to get the temporal profile $\textbf{V}'_{LR}$ corresponding  to the input  $\overline{\textbf{V}}_{LR}$ such that $\textbf{V}'_{LR} = f_r(\overline{\textbf{V}}_{LR})$, where $\textbf{V}'_{HR}$ is the HFR temporal profile of the LFR input $\overline{\textbf{V}}_{LR}$ . We denote the paired data $(\overline{\textbf{V}}_{LR}, \textbf{V}_{HR})$ as $\mathcal{D}_{t}$ . The loss function to update the TSR module $\mathcal{F}_{\theta}$ 
% \amit{I think this is not correct}
is given below.
\begin{align}
    \mathcal{L}^{\mathcal{D}_t} =  \sum\limits_{(\overline{\textbf{V}}_{LR}, \textbf{V}_{HR})\sim\mathcal{D}_t}\Big\Vert \mathcal{F}_{\theta}(\overline{\textbf{V}}_{LR})-\textbf{V}_{HR}\Big\Vert_1
    \label{eqn:L_Dt}
\end{align}
\noindent \textbf{Dynamic Task Generator}  The Dynamic Task Generator (DTG; see Fig.~\ref{fig:overview}a) generates tasks for meta-training on-the-fly using diverse degradation settings. In our approach, the task $\mathcal{T}_i$ is the combination of the spatial down-scaling kernel and temporal sub-sampling method. For spatial down-sampling by a factor of 4 we randomly apply the anisotropic Gaussian kernels using the function $f_s$. Temporal sub-sampling is performed with the function $f_t$ by either selecting alternate frames or by averaging a window of size 3 to obtain a low-frame rate video. \smallskip
% \begin{align}
%     Spatial Down-sample: \textbf{V} =  \sum\limits_{(\textbf{V}'_{HR}, \textbf{V}_{LR})\sim\mathcal{D}_t}\Big\Vert S_{\phi}(\overline{\textbf{V}}_{lr})-\textbf{V}_{LR}\Big\Vert_1
%     \label{eqn:L_Dt}
% \end{align}

\noindent \textbf{Meta-Transfer Learning.} We seek to find a set of transferable initial parameters where a few-gradient steps can adapt the model to the current video and achieve to large performance gain. Motivated by MAML~\cite{finn2017model} and ~\cite{soh2020meta}, we employ meta-transfer learning strategy for spatio-temporal video super-resolution (STVSR) to learn adaptive weights. Unlike MAML, we use the external dataset for meta-training and leverage internal learning for meta-test step. Training with external dataset helps the meta-leaner to focus more on the down-scaling kernel-agnostic property, whereas internal learning helps to exploit the instance specific internal statistics. 

Lines 10-19 in Algorithm~\ref{alg:train} presents the meta-transfer learning optimization protocol.
In our approach, we aim to learn a generalized set of TSR parameters $\theta$ and SSR parameters $\phi$ such that the parameters can adapt to the test video quickly and efficiently in a blind super-resolution setting. The meta-learning achieves this using two steps: task-specific training and blind task adaptation.\smallskip
% To learn generalized set of adaptive parameters, we first sample a task batch $\mathcal{D}_{meta}$ for the task $\mathcal{T}_i$ using Dynamic Task Generator. 
% We then divide $\mathcal{D}_{meta}$ into two subsets: $\mathcal{D}_{tr}$ for task-specific training and $\mathcal{D}_{te}$ for blind task adaption.
% % During training, any $i^{th}$ task $\mathcal{T}_i \in p(\mathcal{T})$ is defined as the combination of spatial and temporal down-sampling applied to the input video. For a given video

% \amit{What are these? Are these videos from the external dataset? Are they tasks in meta-learning? For the latter, I do not know what that means. We need a better definition of task batch.}
% \edits{ Akash: I'll define them below to avoid any confusion. }

\textit{\uline{Task-Specific Training.}} It is the inner loop of the MAML meta-learning algorithm, the meta-learner tries to learn task-specific optimal parameters in one or more gradient descent steps. The inner loop is represented by Lines 12-16 in Algorithm~\ref{alg:train}. Given an external dataset $\mathcal{D}_{HR}$, we obtain a meta-task train batch $\mathcal{D}_{tr} = (\textbf{V}_{LR},{\textbf{V}_s},{\textbf{V}_{ts}})$ for $\mathcal{T}_i \in p(\mathcal{T})$, where $p(\mathcal{T})$ is the task distribution, $\textbf{V}_{LR}$ is LR-LFR video, ${\textbf{V}_s}$ is 4x spatially down-scaled version $\textbf{V}_{LR}$ and ${\textbf{V}_{ts}}$ is 2x temporally down-scaled version of $\textbf{V}_{s}$. We train the TSR model ($\mathcal{F}_{\theta}$) to generate a video $\widehat{\textbf{V}}_s$ with temporal resolution twice to that of input video ($\textbf{V}_{ts}$). The SSR model ($\mathcal{F}_{\phi}$) takes the output of TSR model ($\widehat{\textbf{V}}_s$) and reconstructs the LR-LFR video $\widehat{\textbf{V}}_{LR}$. The output of both the models are given by \vspace{-1mm}
\begin{align}
     \widehat{\textbf{V}}_s = \mathcal{F}_{\theta}(\textbf{V}_{ts}) &&
     \widehat{\textbf{V}}_{LR} = \mathcal{S}_{\phi}(\widehat{\textbf{V}}_s)
    \label{eqn:tr_F}
\end{align}

\input{algorithms/meta_training}
\input{algorithms/meta_testing}

We optimize both the networks for $n_i$ iteration to increase the resolution of a video for a task defined by random spatial and temporal down-scaling kernels. The loss function for the task-specific training is computed as follows:\vspace{-2mm}
\begin{align}
     \mathcal{L}^{tr}_{\mathcal{T}_i} =  \sum\limits_{\mathcal{D}_{tr}} \sum\limits_{n_i} \bigg( \mathcal{L}\Big(\widehat{\textbf{V}}_s, \textbf{V}_s\Big) + \mathcal{L}\Big(\widehat{\textbf{V}}_{LR}, \textbf{V}_{LR}\Big)\bigg)
    \label{eqn:L_tr}
\end{align}
\noindent where $\mathcal{L}$ is reconstruction loss, $n_i$ is number of inner loop iterations for task-specific training. For one gradient update, new adapted parameters $\theta_{i}$ and $\phi_i$ are then obtained as
\begin{align}
\theta_{i} = \theta - \alpha \nabla_{\theta} \mathcal{L}^{tr}_{\mathcal{T}_{i}}(\theta, \phi)\\
\phi{i} = \phi - \alpha \nabla_{\phi} \mathcal{L}^{tr}_{\mathcal{T}_{i}}(\theta, \phi)
\end{align}
where $\alpha$ is the task-level learning rate.

%%%%%%%%%%%%%%%%%%%%%%%%%%%%%%%%%%%%%%%%%%%%%%%%%%%%%%%%%%%%%%%%%%%%%%%%%%%%%%%%%%%%%%
\uline{Blind Task Adaptation.} The blind task-adaptation is the outer loop of meta-learning which adapts the model parameters to the novel task. Here the meta-test batch $\mathcal{D}_{te}$ is sampled from $\mathcal{D}_{HR}$ such that $\mathcal{D}_{te} = (\textbf{V}_{HR},{\widetilde{\textbf{V}}_{LR}},{\textbf{V}_{LR}})$ for $\mathcal{T}_j \in p(\mathcal{T})$ where $\mathcal{T}_i \neq \mathcal{T}_j$, $\textbf{V}_{HR}$ is HR-LFR video, $\widetilde{\textbf{V}}_{LR}$ is 4x spatially down-scaled version of $\textbf{V}_{HR}$ and ${\textbf{V}_{LR}}$ is 2x temporally down-scaled version of $\widetilde{\textbf{V}}_{LR}$.
In order to adapt the models to new task $\mathcal{T}_j$, the model parameters $\theta$ and $\phi$ are optimized to achieve minimal test error on $\mathcal{D}_{te}$ with respect to $\theta_{i}$ and $\phi_{i}$. The meta-objective for blind task-adaptation is
\begin{alignat}{1}
   &\arg \min_{\theta, \phi} \sum_{\mathcal{T}_j\sim p(\mathcal{T})}  \mathcal{L}^{te}_{\mathcal{T}_{j}}(\theta_{i}, \phi_i) \\ \nonumber
    = &\arg \min_{\theta, \phi} \sum_{\mathcal{T}_j} \mathcal{L}^{te}_{\mathcal{T}_{j}}(\theta - \alpha \nabla_{\theta} \mathcal{L}^{tr}_{\mathcal{T}_{i}}, ~~\phi - \alpha \nabla_{\phi} \mathcal{L}^{tr}_{\mathcal{T}_{i}})
\label{eq:L_te}
\end{alignat}
Blind task adaptation using equation~\eqref{eq:L_te} learns the knowledge across tasks $\mathcal{T}_i$ and $\mathcal{T}_j$. The parameter update rule for for the above optimization can be expressed as:
\begin{align}
    \theta \leftarrow \theta - \beta \nabla_{\theta} \sum_{\mathcal{T}_{j}\sim p(\mathcal{T})} \mathcal{L}^{te}_{\mathcal{T}_{j}}(\theta_i,\phi_i) \\
	\phi \leftarrow \phi - \beta \nabla_{\phi} \sum_{\mathcal{T}_{j}\sim p(\mathcal{T})} \mathcal{L}^{te}_{\mathcal{T}_{j}}(\theta_i,\phi_i)
\end{align}
\noindent where $\beta$ is the learning rate for blind task adaptation step.
% \amit{Based on the changes earlier, some of the notation here may need to be changed. Check carefully.}\edits{Akash: Updated the details above}

\subsection{Internal Learning and Inference}
Algorithm~\ref{alg:2} presents the internal learning and inference steps of our proposed approach. Given a LR-LFR video, we spatially down-sample it with corresponding
down-sampling kernel by adopting the kernel estimation algorithms in ~\cite{michaeli2013nonparametric, pan2016blind} for blind scenario and select alternate frames from the LR-LFR video to generate $\textbf{V}_{I}$ and perform a few gradient updates with respect to the model parameter using a single pair of $\textbf{V}_{I}$ as input and a given LR-LFR video $\textbf{V}_{LR}$ as ground truth (Algorithm~\ref{alg:2} Line 2-5). The aim here is to learn the internal statistics of the given video which can be utilized while generating HR-HFR video during inference. The objective function for internal learning is given:
\begin{align}
    \mathcal{L}_{int} =  \Big\Vert \mathcal{S}_{\phi}\big(\mathcal{F}_{\theta}(\textbf{V}_{I})\big)-\textbf{V}_{LR}\Big\Vert_1
    \label{eqn:L_int}
\end{align}
Then, we use the model trained with internal learning for inference. We feed the given LR-LFR input video $\textbf{V}_{LR}$ to the model to generate a HR-HFR video $\widehat{\textbf{V}}_{HR}$ as shown in Algorithm~\ref{alg:2} Line 6.
\label{ssec:int_learn}

%% file: algorithms/meta_training.tex
\begin{algorithm}[t]
    \setstretch{1.05}
	\DontPrintSemicolon
	\caption{\textbf{Ada-VSR} External Training}
    \label{alg:train}
    \KwInput{High-resolution high-frame rate dataset $\mathcal{D}_{HR}$ and task distribution $p(\mathcal{T})$}
    \KwInput{$\alpha, \beta$: task-specific and adaptation learning rate}
    \KwOutput{\textbf{Ada-VSR} model parameters $\theta$ and $\phi$}
    \smallskip
    
    \tcc{Large-Scale training}
    Randomly initialize $\theta$, $\phi$\\
    Generate $\mathcal{D}_s$ using bi-cubic down-sampling
    kernel on $\mathcal{D}_{HR}$ \\ 
	\While{not done}
	{
	    \tcc{Train SSR module}
		Sample LR-HR batch from $\mathcal{D}_s$\\
		Compute $\mathcal{L}^{\mathcal{D}_s}$ by Eq.~\eqref{eqn:L_Ds}\\
		Update $\phi$ with respect to $\mathcal{L}^{\mathcal{D}_s}$ \\
		\tcc{Train TSR module}
		Sample LFR-HFR batch from $\mathcal{D}_t$\\
		Compute $\mathcal{L}^{\mathcal{D}_t}$ by Eq.~\eqref{eqn:L_Dt}\\
		Update $\theta$ with respect to $\mathcal{L}^{\mathcal{D}_t}$
		
	}
	\smallskip
	\tcc{Meta-Transfer Learning}
	\While{not done}
	{
		Sample task batch $\mathcal{D}_{tr}$, $\mathcal{D}_{te}$ for the task $\mathcal{T}_i, \mathcal{T}_j \sim p(\mathcal{T})$\\
		\tcc{Task-Specific Training (inner loop)}
		\For{all $\mathcal{T}_{i}$}{
			Compute meta-training loss ($\mathcal{D}_{tr}$): $\mathcal{L}^{tr}_{\mathcal{T}_{i}}(\theta, \phi)$ \\
			Adapt parameters with gradient descent: \\ \nonumber 
			$\theta_i=\theta - \alpha \nabla_{\theta} \mathcal{L}^{tr}_{\mathcal{T}_{i}}(\theta, \phi)$, \hspace{1em}$\phi_i=\phi - \beta \nabla_{\phi} \mathcal{L}^{tr}_{\mathcal{T}_{i}}(\theta, \phi)$
		}
		\tcc{Blind Task Adaptation (outer loop)}
		Update $\theta$ and $\phi$ with respect to average test loss ($\mathcal{D}_{te}$):\\
		$\theta \leftarrow \theta - \alpha \nabla_{\theta} \sum_{\mathcal{T}_{i}\sim p(\mathcal{T})} \mathcal{L}^{te}_{\mathcal{T}_{i}}(\theta_i,\phi_i)$ \\
		$\phi \leftarrow \phi - \beta \nabla_{\phi} \sum_{\mathcal{T}_{i}\sim p(\mathcal{T})} \mathcal{L}^{te}_{\mathcal{T}_{i}}(\theta_i,\phi_i)$
	}
	
\end{algorithm}

% \begin{algorithm}
% 	\DontPrintSemicolon
% 	\SetAlgoLined
	
% 	\KwInput{LR test image $\mathbf{I}_{LR}$, meta-transfer trained model parameter $\theta_M$, number of gradient updates $n$ and learning rate $\alpha$}
% 	\KwOutput{Super-resolved image $\mathbf{I}_{SR}$}
% 	Initialize model parameter $\theta$ with $\theta_M$\\
% 	Generate LR son $\mathbf{I}_{son}$ by downsampling $\mathbf{I}_{LR}$ with corresponding blur kernel.\\
% 	\For{n steps}
% 	{
% 		Evaluate loss $\mathcal{L}(\theta)=||\mathbf{I}_{LR}-f_\theta (\mathbf{I}_{son})||_1$\\
% 		Update $\theta \leftarrow \theta - \alpha \nabla_{\theta}\mathcal{L}(\theta)$
		
% 	}
% 	\Return $\mathbf{I}_{SR}=f_\theta(\mathbf{I}_{LR})$
	
% 	\caption{Meta-Test}
% 	\label{alg:2}
% \end{algorithm}

%% file: algorithms/meta_testing.tex
\begin{algorithm}[t]
    % \setstretch{0.98}
    \caption{\textbf{Ada-VSR} Internal Learning}
	\label{alg:2}
	\DontPrintSemicolon
	
	\KwInput{LR-LFR test video $\mathbf{V}_{LR}$, meta-transfer trained model parameter $\theta, \phi$, number of gradient updates $n$ and learning rate $\gamma$}
	\KwOutput{High-resolution high-frame rate video $\widehat{\textbf{V}}_{HR}$}
	Generate down-sampled video $\mathbf{V}_{I}$ by down-sampling $\mathbf{V}_{LR}$ with corresponding blur kernel.\\
	\tcc{Internal Learning}
	\For{n steps}
	{
		Evaluate loss $\mathcal{L}_{int}(\theta, \phi)$ using ~\eqref{eqn:L_int}\\
		Update $\theta \leftarrow \theta - \gamma \nabla_{\theta}\mathcal{L}_{int}(\theta, \phi)$ \\ 
		Update $\phi \leftarrow \phi - \gamma \nabla_{\phi}\mathcal{L}_{int}(\theta, \phi)$
	}
	\tcc{Inference to generate HR-HFR}
	\Return $\widehat{\textbf{V}}_{HR}=\mathcal{S}_{\phi}\big(\mathcal{F}_{\theta}(\textbf{V}_{LR})\big)$
\end{algorithm}

%% file: sections/4_experiments.tex
\section{Experiments}

\begin{figure*}[t]
    \centering
    \includegraphics[width=0.995\linewidth]{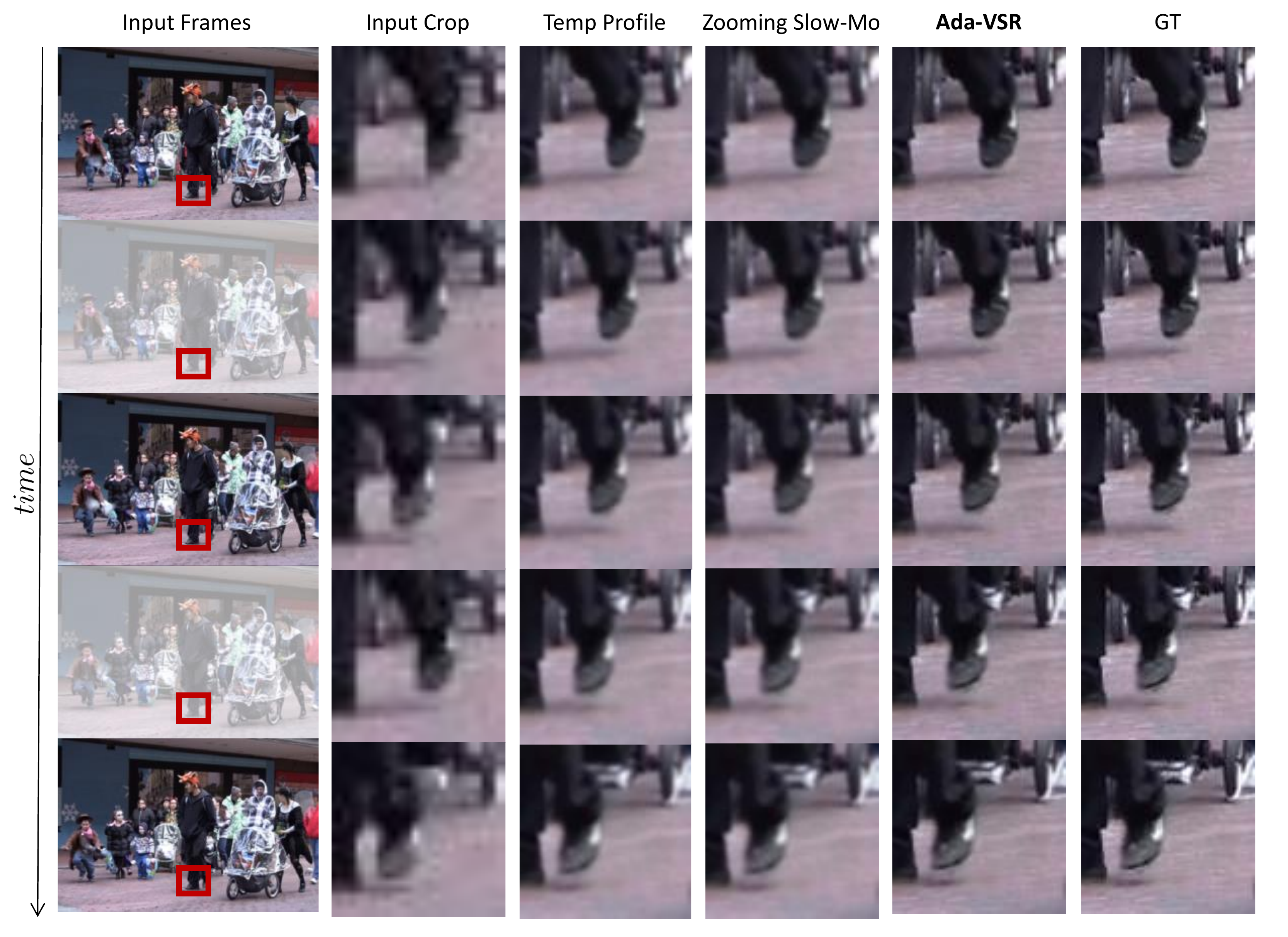}
    % \vspace{-2mm}
    \caption{Comparison of qualitative results with the state-of-the-art methods. %Input frames are obtained by selecting alternate frames of the HR-HFR video followed by  down-sampling using random kernel other than bi-cubic. 
    First column consists of the low-resolution low-frame-rate video input with  non-bi-cubic degradation and the missing frames are faded. Second column and last column are the input and ground-truth crop of the input frame region marked in red. As opposed to temporal profile approach, Zooming Slow-Mo and Ada-VSR perform better as they can exploit internal structure of the input. Ada-VSR produces visually appealing results than Zooming Slow-Mo as the weights can easily adapt to novel tasks. Zoom-in for better visualization.}
    \label{fig:qual}
\end{figure*}

\input{tables/sota}

In this section, we first introduce the benchmark datasets and evaluation metrics. The qualitative and quantitative experiments are shown to demonstrate the effectiveness of our proposed approach in generating high-resolution high frame-rate videos.

\subsection{Datasets and Metrics}
We evaluate the performance of our approach using publicly available Vimeo-90K~\cite{xue2019video} and Vid4~\cite{liu2011bayesian} datasets which have been used in many prior spatial video super-resolution and temporal super-resolution works.\smallskip

\noindent \textbf{Vimeo-90K Dataset.} The Vimeo-90K~\cite{xue2019video} dataset contains 91,707 short video clips, each containing 7 frames. The spatial resolution of each frame is ($448 \times 256$). We use Vimeo-90K only for pre-training and meta-training, using the training split of 64,612 clips and use to the test set to compare against state-of-the-art approaches.\smallskip

\noindent \textbf{Vid4 Dataset.} The Vid4 dataset~\cite{liu2011bayesian},contains four video
sequences: city, walk, calendar, and foliage. All the videos in Vid4 dataset contain at least 30 frames each and are of spatial resolution 720×480. We evaluate a model trained on Vimeo-90K dataset on the Vid4 dataset and report the performance.\smallskip

% \noindent \textbf{Dataset Preparation.} For large-scale training with Vimeo90K~\cite{xue2019video} dataset, the 4-times bi-cubic down-sampled odd-indexed frames are considered as LR-LFR input, and the corresponding consecutive frames are considered as HF-HFR ground truth. The input and ground truth frames are divided into patches of $64 \times 64$ for training the network.\smallskip

\noindent \textbf{Metrics.} For quantitative evaluation, we compare three metrics that evaluate different aspects of output image quality: Peak Signal-to-Noise Ratio (PSNR) \cite{huynh2008scope}, Structural Similarity Index Measure (SSIM) \cite{wang2004image} and Naturalness Image Quality Evaluator (NIQE) \cite{mittal2012making}. 

%PSNR uses ground truth image as reference to measure of the extent of noise and distortion caused due to image compression. For the task at hand, PSNR measures the extent of distortion free super-resolution achieved. Higher the PSNR (in dB), the better the desired performance. SSIM, like PSNR, is a full reference metric. While PSNR calculates the absolute value of degradation, SSIM calculates perceived quality by combining image properties such as local structure, luminance and texture. This aligns better with human perceived quality and the higher the SSIM, the better the performance. NIQE, on the other hand, is a no reference metric. A NIQE model is computed on a predetermined dataset and the quality index is calculated using only output image (no reference). This measures the overall quality and naturalness of an image with arbitrary distortions. A lower value of NIQE is desirable. 

\subsection{Implementation Details}
Our framework is implemented in PyTorch~\cite{paszke2017automatic}. All the experiments are trained with a batch size of 32.  We employ ADAM optimizer as the meta-optimizer in the meta-transfer learning step. The task-specific learning rate $\alpha$ is set to 0.01 and the adaptation learning rate $\beta$ is set to 0.0001 for all our training experiments. The number of iterations in the task-specific training $n_i$ is set to 10. We extracted training patches with a size of $64\times64$ for large-scale training. We utilize the Vimeo-90K dataset train split as the external dataset for large-scale training and meta-transfer learning. For internal learning the learning rate $\gamma$ is set to 0.0001.
% \subsection{Results and Ablation}
\subsection{Qualitative Results.}
Figure~\ref{fig:qual} compares a high-resolution high frame-rate videos generated using the proposed \textbf{Ada-VSR} approach with other state-of-the-art methods given a low-resolution low frame-rate video (left column). The low-resolution low-frame-rate input video is obtained by applying a  non-bicubic degradation to the alternate frame of a high-resolution high-frame rate video. The skipped frames are faded in the Figure~\ref{fig:qual}.
It can be seen that the temporal profile based approach~\cite{xiao2020space} does not perform well. It is due the fact that the model is trained with the assumption that there is a bi-cubic relationship between the LR-LFR and HR-HFR videos. This assumption is violated when we use an input video which was obtained by a non-bicubic down-sampling. Zooming Slow-Mo~\cite{xiang2020zooming} produces slightly better video compared to the temporal profile approach as it is able to exploit the internal structure within the video. However, it is not able to exploit the external knowledge and the output is still blurry. Our proposed approach, \textbf{Ada-VSR}, produces higher-quality and more visually appealing output as compared to both of these approaches. The performance of our approach can be attributed to the adaptive parameters learned on external dataset with meta-learning. These parameters provide good initial parameters for internal training to learn instance specific characteristics.

\subsection{Quantitative Results}
Our proposed method performs joint spatio-temporal video super-resolution. We compare \textbf{Ada-VSR} against representative STVSR approaches. We first compare our work with the two-stage solutions by cascading a  temporal super-resolution module (TSR) and a spatial super-resolution module (SSR).
%\amit{check this sentence - rewritten}
For temporal super-resolution module (TSR) we select SepConv~\cite{niklaus2017videosepconv} and DAIN~\cite{bao2019depth} model, while SAN~\cite{dai2019san}, IMDN~\cite{hui2019imdn}, DynaVSR~\cite{lee2021dynavsr}, and EDVR~\cite{wang2019edvr} are selected for spatial video super-resolution module (SSR). We also compare our work with recently proposed one-stage STVSR Zooming Slow-Mo~\cite{xiang2020zooming} and Temporal profile based approach~\cite{xiao2020space} where spatio-temporal super-resolution is performed jointly.

Table~\ref{tab:sota} presents the quantitative comparison on the test-set of Vimeo-90K~\cite{xue2019video} dataset and Vid4~\cite{liu2011bayesian} dataset. It is evident that our approach outperforms all the two-stage approaches by a significant margin against all the three metrics. When compared with the state-of-the-art one-stage approaches, temporal profile based~\cite{xiao2020space} and Zooming Slow-Mo~\cite{xiang2020zooming}, our proposed \textbf{Ada-VSR} achieves superior performance on both the dataset except on Vimeo-90K Slow dataset in terms of PSNR where our performance is only 0.04db less than the temporal profile based approach.

In Table~\ref{tab:time_analysis}, we compare the average inference time of different state-of-the-art approaches. As our method learns adaptive weights that can easily adapt to novel tasks, only a few gradient updates during internal learning step are required to achieve visually compelling results. It can be observed from Table~\ref{tab:time_analysis} that we outperform the Zooming Slow-Mo~\cite{xiang2020zooming} by a margin of $0.66$dB and the temporal profile approach by a significant margin of $+1.18$dB. As mentioned earlier, the temporal profile approach assumes that the degradation is bi-cubic hence it cannot generalize well on videos degraded using different blur kernels. It should also be noted that our approach is at about twice as fast as the temporal profile approach and at least 3 times faster when compared with Zooming Slow-Mo. Thus, significantly reducing the inference time during for new test videos with unknown degradation. Values for Zooming Slow-Mo~\cite{xiang2020zooming} and temporal-profile~\cite{xiao2020space} are reported from ~\cite{xiao2020space}.

\input{tables/time_analysis}
\input{tables/ablation}

\subsection{Ablation Study} We investigate the contribution of large-scale training of the TSR module ($\mathcal{F}_{\theta}$) and the SSR module ($\mathcal{S}_{\phi}$). First, for the task-specific training in meta-learning, only the SSR module is initialized using the pre-trained weights obtained in the large-scale training step and the TSR module is initialized randomly. In second case, only the TSR module is initialized with the weights obtained from the large-scale training and the SSR module is randomly initialized.
The quantitative results of impact of large-scale training using external data on Vid4 dataset are shown in Table~\ref{tab:ablation}. We can observe that the performance of \textbf{Ada-VSR} drops if large-scale pre-training is performed on only one of the SSR or TSR module. This is expected since the meta-learning algorithm MAML~\cite{finn2017model} has shown to be unstable when training without a warm model initialization. It is interesting to note that the model with large-scale training of only TSR module outperforms the one initialized with only the SSR module. We believe this could be due to the rich information available in temporal profiles used for large-scale training of the TSR module which provide stable initial parameters for task-specific training and blind-task adaptation during meta-training.

%% file: tables/sota.tex
\begin{table*}[t]
 \caption{{Quantitative results comparison on} Vimeo-90K~\cite{xue2019video} and Vid4~\cite{liu2011bayesian} datasets.  Our proposed approach is very competitive against various state-of-the-art approaches. Best scores are shown in bold and the second best are underlined. The state-of-the-art results are reported from~\cite{xiao2020space}.}
%  \vspace{-1mm}
 \resizebox{0.98\linewidth}{!}{%
\renewcommand{\arraystretch}{1.22}
\begin{tabular}{>{\centering\arraybackslash}p{1.6cm}|>{\centering\arraybackslash}p{1.6cm}|ccc|ccc|ccc|ccc}
\toprule[1.2pt]
\multicolumn{2}{c|}{{\textit{\textbf{Method}}}}                         & \multicolumn{3}{c|}{{\textit{\textbf{Vimeo-90K Slow}}}}                                                          & \multicolumn{3}{c|}{{\textit{\textbf{Vimeo-90K Medium}}}}                                                       & \multicolumn{3}{c|}{{\textit{\textbf{Vimeo-90K Fast}}}}                                                         & \multicolumn{3}{c}{{\textit{\textbf{Vid4}}}}                                                                  \\ \toprule[1.2pt]
\textbf{TSR} & \textbf{SSR} & \textbf{PSNR} $\uparrow$ & \textbf{SSIM} $\uparrow$ & \textbf{NIQE} $\downarrow$ & \textbf{PSNR} $\uparrow$ & \textbf{SSIM} $\uparrow$ & \textbf{NIQE} $\downarrow$ & \textbf{PSNR} $\uparrow$ & \textbf{SSIM} $\uparrow$ & \textbf{NIQE} $\downarrow$ & \textbf{PSNR} $\uparrow$ & \textbf{SSIM} $\uparrow$ & \textbf{NIQE} $\downarrow$ \\
\toprule[1.2pt]
\multicolumn{1}{c|}{SepConv~\cite{niklaus2017videosepconv}}      & {IMDN~\cite{hui2019imdn}}         & {31.75}         & {0.88}          & {7.68}          & {33.13}         & {0.90}          & {7.78}          & {34.31}         & {0.92}          & {8.55}          & {24.87}         & {0.72}          & {6.34}          \\ \hline
\multicolumn{1}{c|}{SepConv~\cite{niklaus2017videosepconv}}      & {SAN~\cite{dai2019san}}          & {32.12}         & {0.90}          & {7.10}          & {33.59}         & {0.91}          & {7.46}          & {34.97}         & {0.92}          & {8.48}          & {24.93}         & {0.72}          & {5.89}          \\ \hline
\multicolumn{1}{c|}{SepConv~\cite{niklaus2017videosepconv}}      & {EDVR~\cite{wang2019edvr}}         & {32.97}         & {0.91}          & {7.00}          & {34.25}         & {0.92}          & {7.40}          & {35.51}         & {0.92}          & {8.48}          & {25.93}         & {0.78}          & {5.70}          \\ \hline
\multicolumn{1}{c|}{DAIN~\cite{bao2019depth}}         & {IMDN~\cite{hui2019imdn}}         & {31.84}         & {0.89}          & {7.13}          & {33.39}         & {0.91}          & {7.58}          & {34.74}         & {0.92}          & {8.43}          & {24.93}         & {0.72}          & {6.18}          \\ \hline
\multicolumn{1}{c|}{DAIN~\cite{bao2019depth}}         & {SAN~\cite{dai2019san}}          & {32.26}         & {0.90}          & {7.05}          & {33.82}         & {0.92}          & {7.45}          & {35.27}         & {0.92}          & {8.48}          & {25.14}         & {0.73}          & {5.78}          \\ \hline
\multicolumn{1}{c|}{DAIN~\cite{bao2019depth}}         & {EDVR~\cite{wang2019edvr}}         & {33.21}         & {0.91}          & {7.06}          & {34.73}         & {0.93}          & {7.39}          & {35.71}         & {0.93}          & {8.47}          & {26.12}         & {0.79}          & {5.62}          \\\hline\hline \toprule[1.2pt]
\multicolumn{1}{c|}{\textbf{SSR}} & {\textbf{TSR}} & \textbf{PSNR} $\uparrow$ & \textbf{SSIM} $\uparrow$ & \textbf{NIQE} $\downarrow$ & \textbf{PSNR} $\uparrow$ & \textbf{SSIM} $\uparrow$ & \textbf{NIQE} $\downarrow$ & \textbf{PSNR} $\uparrow$ & \textbf{SSIM} $\uparrow$ & \textbf{NIQE} $\downarrow$ & \textbf{PSNR} $\uparrow$ & \textbf{SSIM} $\uparrow$ & \textbf{NIQE} $\downarrow$ \\ \toprule[1.2pt]
\multicolumn{1}{c|}{IMDN~\cite{hui2019imdn}}         & {SepConv~\cite{niklaus2017videosepconv}}      & {32.01}         & {0.89}          & {7.67}          & {33.22}         & {0.90}          & {7.65}          & {34.50}         & {0.92}          & {8.54}          & {24.88}         & {0.72}          & {6.33}          \\ \hline
\multicolumn{1}{c|}{IMDN~\cite{hui2019imdn}}         & {DAIN~\cite{bao2019depth}}         & {32.27}         & {0.89}          & {6.99}          & {33.73}         & {0.92}          & {7.17}          & {35.15}         & {0.92}          & {8.41}          & {24.99}         & {0.72}          & {6.2}           \\ \hline
\multicolumn{1}{c|}{SAN~\cite{dai2019san}}          & {SepConv~\cite{niklaus2017videosepconv}}      & {32.32}         & {0.90}          & {6.99}          & {33.73}         & {0.92}          & {7.32}          & {35.33}         & {0.92}          & {8.42}          & {25.01}         & {0.73}          & {5.87}          \\ \hline
\multicolumn{1}{c|}{SAN~\cite{dai2019san}}          & {DAIN~\cite{bao2019depth}}         & {32.56}         & {0.91}          & {6.90}          & {34.12}         & {0.93}          & {7.43}          & {35.47}         & {0.92}          & {8.39}          & {25.26}         & {0.75}          & {6.16}          \\
\hline
\multicolumn{1}{c|}{DynaVSR~\cite{lee2021dynavsr}} &{DAIN~\cite{bao2019depth}}                                            &  -                                  &  -                                  &   -                                 & -                                   &  -                                  &    -                                &    -                                &    -                                &     -                               &  26.54                                  & 0.81                                   &   5.65                                 \\
\hline\hline\toprule[1.2pt]
\multicolumn{2}{c|}{\textbf{End-to-end Framework}}                    & \textbf{PSNR} $\uparrow$ & \textbf{SSIM} $\uparrow$ & \textbf{NIQE} $\downarrow$ & \textbf{PSNR} $\uparrow$ & \textbf{SSIM} $\uparrow$ & \textbf{NIQE} $\downarrow$ & \textbf{PSNR} $\uparrow$ & \textbf{SSIM} $\uparrow$ & \textbf{NIQE} $\downarrow$ & \textbf{PSNR} $\uparrow$ & \textbf{SSIM} $\uparrow$ & \textbf{NIQE} $\downarrow$ \\
\toprule[1.2pt]

\multicolumn{2}{l|}{Zooming Slow-Mo~\cite{xiang2020zooming}}                                  & {33.29}         & {0.91}          & {6.94}          & {35.24}         & {0.93}          & {7.35}          & \uline{36.43}         & {0.93}          & {8.41}          & {26.30}         & {0.80}          & {5.62}          \\ \hline
\multicolumn{2}{l|}{Temporal Profile~\cite{xiao2020space}}                                             & \textbf{33.40}                              & \textbf{0.92}                               & \uline{6.17}                               & \uline{35.55}                              & \uline{0.94}                               & \uline{6.37}                               & 36.29                              & \uline{0.93}                               & \uline{7.13}                               & \uline{26.50}                              & \uline{0.82}                               & \uline{5.48}                               \\
\hline
\multicolumn{2}{l|}{\textbf{Ada-VSR (Ours)}}                                     &  \uline{33.36}                                  &  \textbf{0.92}                                  &   \textbf{6.12}                                 &   \textbf{35.91}                                 &   \textbf{0.95}                                 & \textbf{6.33}                                    &    \textbf{36.52}                                & \textbf{0.95}                                   &    \textbf{6.99}                                & \textbf{26.98}                                   &   \textbf{0.84}                                 & \textbf{5.40}\\                                 
\toprule[1.2pt]
\end{tabular}
}
\label{tab:sota}
\end{table*}

%% file: tables/time_analysis.tex
\begin{table}[t]
\caption{Average inference time (sec per frames) comparison of Ada-VSR with recent approaches for blind spatio-temporal video super-resolution.}
\resizebox{0.98\columnwidth}{!}{%
\renewcommand{\arraystretch}{1.22}
\label{tab:time_analysis}

\begin{tabular}{>{\raggedright\arraybackslash}p{3.5cm}|>{\centering\arraybackslash}p{1.7cm}|>{\centering\arraybackslash}p{1.7cm}}
\toprule[1.2pt]
\multicolumn{1}{c|}{\multirow{2}{*}{\textit{\textbf{Method}}}}                            & \multicolumn{2}{c}{\textit{\textbf{Vid4~\cite{liu2011bayesian}}}} \\ \cline{2-3} 
\multicolumn{1}{c|}{}                                                            & \multicolumn{1}{c|}{\textbf{PSNR}$\uparrow$}        & \textbf{Avg. time}$\downarrow$       \\ \toprule[1.2pt]

Zooming Slow-Mo~\cite{xiang2020zooming}                                     & 26.30                                     & 0.1995                                           \\ \hline
DynaVSR~\cite{lee2021dynavsr} + DAIN~\cite{bao2019depth} & {\ul 26.54}                               & 0.8940                                           \\ \hline
Temporal profile~\cite{xiao2020space}                                    & 25.78                                     & {\ul 0.1328}                                     \\ \hline
\textbf{Ada-VSR (Ours)}                                                           & \textbf{26.96}                            & \textbf{0.0680}                                  \\
\toprule[1.2pt]
\end{tabular}
}
\end{table}

%% file: tables/ablation.tex
\begin{table}[t]
\caption{Impact of large-scale training of TSR and SSR modules in Ada-VSR on the target performance.}
\resizebox{0.98\columnwidth}{!}{%
\renewcommand{\arraystretch}{1.22}
\label{tab:ablation}
\begin{tabular}{M{1.3cm}|M{1.3cm}|M{1.1cm}|M{1.1cm}|M{1.1cm}}
\toprule[1.2pt]
\multicolumn{2}{c|}{\textit{\textbf{External-Training}}} & \multicolumn{3}{c}{\textit{\textbf{Vid4~\cite{liu2011bayesian}}}}\\
\cline{1-5}
\textbf{Spatial} \quad & \textbf{Temporal} & \textbf{PSNR} $\uparrow$ & \textbf{SSIM}$\uparrow$ & \textbf{NIQE}$\downarrow$\\
\toprule[1.2pt]
% CNN~\cite{kappeler2016video}         & \textcolor{ao(english)}{\boldcheckmark}   & \textcolor{cadmiumred}{\xmark}            & \textcolor{cadmiumred}{\xmark}          & \textcolor{cadmiumred}{\xmark}\\
\hline
\textcolor{ao(english)}{\boldcheckmark}  & \textcolor{cadmiumred}{\xmark}  & 25.98 & 0.80 & 5.77   \\
\hline
\textcolor{cadmiumred}{\xmark}   & \textcolor{ao(english)}{\boldcheckmark}   & 26.27 & 0.81 & 5.59 \\
\hline
\textcolor{ao(english)}{\boldcheckmark}   & \textcolor{ao(english)}{\boldcheckmark}   & \textbf{26.98}  & \textbf{0.84} & \textbf{5.40} \\
\bottomrule[1.2pt]

\end{tabular}
}
\end{table}

%% file: sections/5_conclusion.tex
\section{Conclusions}

We present an Adaptive Video Super Resolution framework (\textbf{Ada-VSR}) for generating high resolution high frame-rate videos from low resolution low frame-rate input videos. We leverage external as well as internal learning to achieve spatio-temporal super-resolution. Specifically, external learning employs meta-learning to learn adaptive network parameters that can easily adapt unknown degradation, while internal learning, on the other hand, helps to capture the underlying statistics of down-sampling and degradation specific to the input video by exploiting the internal structure, thereby making our approach more suited for practical, real-world data enhancement tasks. The proposed approach is able to achieve superior enhancement while adapting to unknown degradation models as shown in our experiments. Experiments on standard datasets show not only the quantitative and qualitative efficacy of our proposed model in joint spatio-temporal video super-resolution, but also the improvement in computational time over various state-of-the-art methods.

% We compare quantitative metrics against various methods and perceptual results using top 2 methods where we demonstrate that Ada-VSR is able to achieve superior enhancement while adapting to unknown degradation models. 
% Additionally, leveraging the internal learning in meta-testing allows us to achieve superior performance by reducing the gradient steps - lowering the inference time significantly. 

%% file: sections/6_acknowledgments.tex
\begin{acks}
The work was partially supported by NSF grants 1664172, 1724341 and 1911197.
\end{acks}